\documentstyle[prl,aps,tighten]{revtex}
\begin{document}
\twocolumn[
\hsize\textwidth\columnwidth\hsize\csname@twocolumnfalse\endcsname

\title{Chaos in Superstring Cosmology}
\author{Thibault Damour}
\address{Institut des Hautes Etudes Scientifiques, 35, route de 
Chartres, F-91440 Bures-sur-Yvette, France \\ 
and \\
Institute for Theoretical Physics, University of California, Santa 
Barbara, CA 93106-4030, USA}
\author{Marc Henneaux}
\address{Physique Th\'eorique et Math\'ematique, Universit\'e Libre 
de Bruxelles, C.P. 231, B-1050, Bruxelles, Belgium \\
and \\
Centro de Estudios Cient\'{\i}ficos, Casilla 1469, Valdivia, Chile}

\maketitle

\begin{abstract}
It is shown that the general solution near a spacelike
singularity of the Einstein-dilaton-$p$-form field equations 
relevant to superstring theories and $M$-theory 
exhibits an oscillatory behaviour of the 
Belinskii-Khalatnikov-Lifshitz type.
String dualities play a significant role in the analysis.
\end{abstract}

]

\vglue 2cm

An outstanding result in theoretical cosmology has been the 
discovery by Belinskii, Khalatnikov and Lifshitz (BKL) that the 
generic solution of the four-dimensional Einstein's vacuum equations 
near a cosmological singularity exhibits a never ending oscillatory 
behaviour \cite{BKL}. The oscillatory approach toward the 
singularity has the character of a random process, whose 
chaotic nature has been intensively studied \cite{LLK}. 
However, two results cast a doubt on the physical 
applicability, to our universe, of the BKL picture. First, it was 
surprisingly found that the chaotic BKL oscillatory behaviour 
disappears from the generic solution of the vacuum Einstein 
equations in spacetime dimension $D \geq 11$ and is replaced by a 
monotonic Kasner-like power-law behaviour \cite{DHS}. Second, it was 
proved that the generic solution of the four-dimensional 
Einstein-scalar equations also exhibits a non-oscillatory, 
power-law behaviour \cite{BK73}, \cite{AR00}.

Superstring theory \cite{polchinski} suggests that the massless 
(bosonic) degrees of freedom which can be generically excited near a 
cosmological singularity correspond to a high-dimension ($D = 10$ or 
$11$) Kaluza-Klein-type model containing, in addition to Einstein's 
$D$-dimensional gravity, several other fields (scalars, vectors 
and/or forms). In view of the results quoted above, it is a priori 
unclear whether the full field content of superstring theory will 
imply, as generic cosmological solution, a chaotic BKL-like 
behaviour, or a monotonic Kasner-like one. In this letter we  
report the result that the massless bosonic content of all 
superstring models ($D = 10$ IIA, IIB, I, ${\rm het}_{\rm E}$, ${\rm 
het}_{\rm SO}$), as well as of 
$M$-theory ($D=11$ supergravity), generically implies a chaotic 
BKL-like oscillatory behaviour near
a cosmological singularity. 
[Our analysis applies at
scales large enough to excite all Kaluza-Klein-type modes, but small
enough to be able to neglect the stringy and non-perturbative
massive states.]
It is
the presence of various form fields (e.g. the three form in ${\rm 
SUGRA}_{11}$) which provides the crucial source of this generic 
oscillatory behaviour.

Let us consider a model of the general form
\begin{eqnarray}
S &=& \int \, d^D \, x \, \sqrt{g} \big[ R(g) - \partial_{\mu} \, 
\varphi \, \partial^{\mu} \, \varphi \nonumber \\
& & \; \; \; \; \; \; \; \; \; \;
- \sum_p \ \frac{1}{(p+1)!} \ 
e^{\lambda_p \, \varphi} \, (d \, A_p)^2 \big] \, . \label{eq1}
\end{eqnarray}
Here, the spacetime dimension $D$ is left unspecified. We work (as a 
convenient common formulation) in the Einstein conformal frame, and 
we normalize the kinetic term of the ``dilaton'' $\varphi$ with a 
weight 1 with respect to the Ricci scalar. The integer $p \geq 0$ 
labels the various $p$-forms $A_p \equiv A_{\mu_1 \ldots \mu_p}$ 
present in the theory, with field strengths $F_{p+1} \equiv d \, 
A_p$, i.e. $F_{\mu_0 \, \mu_1 \ldots \mu_p} = \partial_{\mu_0} \, 
A_{\mu_1 \ldots \mu_p} \pm p$ permutations. The real parameter 
$\lambda_p$ plays the crucial role of measuring the strength of the 
coupling of the dilaton to the $p$-form $A_p$ (in the Einstein 
frame). When $p=0$,
we assume that $\lambda_0 \not= 0$
(this is the case in type IIB where there is
a second scalar).  
The Einstein metric $g_{\mu \nu}$ is used to lower or raise 
all indices in Eq.~(\ref{eq1}) ($g \equiv -\det \, g_{\mu \nu}$). 
The model (\ref{eq1}) is, as it reads, not quite general enough to 
represent in detail all the superstring actions. Indeed, it lacks 
additional terms involving possible couplings between the form 
fields (e.g. Yang-Mills couplings for $p=1$ multiplets, Chern-Simons 
terms, $(d \, C_2 - C_0 \, d \, B_2)^2$-type terms 
in type IIB). However, we have verified in all relevant 
cases that these additional terms do not qualitatively modify the 
BKL behaviour to be discussed below. On the other hand, in the case 
of $M$-theory (i.e. $D=11$ SUGRA) the dilaton 
$\varphi$ is absent, and one must cancell 
its contributions to the dynamics.

The leading Kasner-like approximation to the solution of the field 
equations for $g_{\mu \nu}$ and $\varphi$
derived from (\ref{eq1}) is, as usual \cite{BKL}, 
\begin{mathletters}
\label{eq2}
\begin{eqnarray}
g_{\mu \nu} \, dx^{\mu} \, dx^{\nu} &\simeq &-dt^2 + \sum_{i=1}^d \, 
t^{2 p_i (x)} \, (\omega^i)^2 \, , \label{eq2a} \\
\nonumber \\
\varphi &\simeq &p_{\varphi} \, (x) \, \ln \, t \, +
\psi \, (x) \, , \label{eq2b} 
\end{eqnarray}
\end{mathletters}
where $d \equiv D-1$ denotes the spatial dimension and where 
$\omega^i \, (x) = e_j^i \, (x) \, dx^j$ is a time-independent 
$d$-bein. The spatially dependent Kasner exponents $p_i \, (x)$, 
$p_{\varphi} \, (x)$ must satisfy the famous Kasner constraints 
(modified by the presence of the dilaton):
\begin{mathletters}
\label{eq3}
\begin{eqnarray}
Q(p) \equiv p_{\varphi}^2 + \sum_{i=1}^d \ p_i^2 - \left( 
\sum_{i=1}^d \ p_i \right)^2 &= &0 \, , \label{eq3a} \\
\nonumber \\
\sum_{i=1}^d \ p_i &= &1 \, . \label{eq3b} 
\end{eqnarray}
\end{mathletters}
The set of parameters satisfying Eqs.~(\ref{eq3}) is topologically
a $(d-1)$-dimensional sphere: the ``Kasner sphere''. When the 
dilaton is absent, one must set $p_{\varphi}$ to zero in 
Eq.(\ref{eq3a}). In that case the dimension of the Kasner sphere is 
$d-2 = D-3$.

The approximate solution Eqs.~(\ref{eq2}) is obtained by neglecting 
in the field equations for $g_{\mu \nu}$ and $\varphi$: (i) the 
effect of the spatial derivatives of $g_{\mu \nu}$ and $\varphi$, 
and (ii) the contributions of the various 
$p$-form fields $A_p$. The condition for the ``stability'' of the 
solution (\ref{eq2}), i.e. for the absence of BKL oscillations at $t 
\rightarrow 0$, is that the inclusion in the field equations of the 
discarded contributions (i) and (ii) (computed within the assumption 
(\ref{eq2})) be fractionally negligible as $t \rightarrow 0$. As 
usual, the fractional effect of the spatial derivatives of $\varphi$ 
is found to be negligible, while the fractional effect (with respect 
to the leading terms, which are $\propto t^{-2}$) of the spatial 
derivatives of the metric, i.e. the quantities $t^2 \, 
\overline{R}_j^i$ (where $\overline{R}_j^i$ denotes the 
$d$-dimensional Ricci tensor) contains, as only ``dangerous terms'' 
when $t \rightarrow 0$ a sum of terms $\propto t^{2 g_{ijk}}$, where 
the {\it gravitational exponents} $g_{ijk}$ ($i \ne j$, $i \ne k$, 
$j \ne k$) are the following combinations of the Kasner exponents 
\cite{DHS}
\begin{equation}
g_{ijk} \, (p) = 2 \, p_i + \sum_{\ell \ne i,j,k} \ p_{\ell} = 1 + 
p_i - p_j - p_k \, . \label{eq4}
\end{equation}
The ``gravitational'' stability condition is that all the exponents 
$g_{ijk} \, (p)$ be positive. In the presence of form fields $A_p$ there 
are additional stability conditions related to the contributions
of the form fields to the Einstein-dilaton equations.
They are obtained by solving, 
\`a la BKL, the $p$-form field equations in the background 
(\ref{eq2}) and then estimating the corresponding ``dangerous'' 
terms in the Einstein field equations. When
neglecting the spatial derivatives in the Maxwell equations in 
first-order form $d \, F = 0$ and $\delta \,
(e^{\lambda_p \, \varphi} \, F) = 0$, where $\delta \equiv * \, d \,
*$ is the (Hodge) dual of the Cartan differential $d$
and $F_{p+1} = d \, A_p$, one 
finds that both the ``electric'' components $\sqrt g \, e^{\lambda_p 
\, \varphi} \, F^{0 i_1 \ldots i_p}$, and the ``magnetic'' 
components $F_{j_1 \ldots j_{p+1}}$, are constant in time. 
Combining this information with the approximate results 
(\ref{eq2}) one can estimate the fractional effect of the $p$-form 
contributions in the right-hand-side of the $g_{\mu \nu}$- and 
$\varphi$-field equations, i.e. the quantities $t^2 \, T_{(A)0}^0$ 
and $t^2 \, T_{(A)j}^i$ where $T_{(A)\nu}^{\mu}$ denotes the 
stress-energy tensor of the $p$-form. [As usual \cite{BKL} the mixed 
terms $T_{(A)i}^0$, which enter the momentum constraints play a 
rather different role and do not need to be explicitly 
considered.] Finally, one gets as ``dangerous'' terms when $t 
\rightarrow 0$ a sum of ``electric'' contributions $\propto \, t^{2 
\, e_{i_1 \ldots i_p}^{(p)}}$ and of ``magnetic'' ones $\propto \, 
t^{2 \, b_{j_1 \ldots j_{d-p-1}}^{(p)}}$. Here, the {\it electric 
exponents} $e_{i_1 \ldots i_p}^{(p)}$ (where all the indices $i_n$ 
are different) are defined as
\begin{equation}
e_{i_1 \ldots i_p}^{(p)} \, (p) = p_{i_1} + p_{i_2} + \cdots + 
p_{i_p} - \frac{1}{2} \ \lambda_p \, p_{\varphi} \, , \label{eq5}
\end{equation}
while the {\it magnetic exponents} $b_{j_1 \ldots j_{d-p-1}}^{(p)}$ 
(where all the indices $j_n$ are different) are
\begin{equation}
b_{j_1 \ldots j_{d-p-1}}^{(p)} \, (p) = p_{j_1} + p_{j_2} + \cdots + 
p_{j_{d-p-1}} + \frac{1}{2} \, \lambda_p \, p_{\varphi} \, . 
\label{eq6}
\end{equation}
To each $p$-form is associated a (duality-invariant)
double family of ``stability'' 
exponents $e^{(p)}$, $b^{(p)}$. The ``electric'' (respectively
``magnetic'') stability condition is that all the exponents $e^{(p)}$
(respectively, $b^{(p)}$) be positive. This result generalizes the results 
of \cite{BK81} on the effect of vector fields in $D=4$.

The main result reported here is that, for all superstring models, 
there exists no open region of the Kasner sphere where all the 
stability exponents $g(p)$, $e(p)$, $b(p)$ are strictly positive. 
To define the set of stability conditions for the various superstring
models, let us review their field content and give the values of the
crucial dilaton couplings $\lambda_p$.
The simplest case is the 
massless bosonic sector of $M$-theory, i.e of SUGRA in $D = 11$. In 
that case, there is a 3-form and no dilaton. 
The parameters $p_{\alpha}^M$, 
$\alpha = 1 , \ldots , 10$, run over the 8-dimensional sphere
$S_M^8$ defined by $\sum_{\alpha} \ (p_{\alpha}^M)^2 
= 1 = \sum_{\alpha} \ p_{\alpha}^M$. 
The presence of a
3-form $A_3$ uncoupled to any dilaton implies that the electric
and magnetic stability exponents are respectively given by (\ref{eq5})
and (\ref{eq6}) with $p=3$, $\lambda_p = 0$ and $d=10$, i.e.,
$e_{\alpha_1 \, \alpha_2 \, \alpha_3}^{M(3)} = p_{\alpha_1}^M + 
p_{\alpha_2}^M + p_{\alpha_3}^M $ and
$b_{\alpha_1 \ldots \alpha_6}^{M(3)} = p_{\alpha_1}^M + \cdots + 
p_{\alpha_6}^M $.

The $D=10$ type IIA string theory involves, besides $g_{\mu 
\nu}$ and a dilaton $\varphi = \Phi / \sqrt 2$ (with $g_s = 
e^{\Phi}$ being the string coupling) a 1-form, a 2-form and a 
3-form. The (Einstein-frame) dilaton coupling parameters of the 
forms are $\lambda_1^A = 3 \, \sqrt 2 / 2$, $\lambda_2^A 
= - \sqrt 2$ and $\lambda_3^A = \sqrt 2 / 2$, respectively. Besides the 
dilaton Kasner exponent $p_{\varphi}^A$, there are nine metric 
exponents $p_i^A$, $i=1,\ldots ,9$. They run over $S_A^8$
defined by Eqs.~(\ref{eq3}). 

The $D=10$ type IIB string theory involves (besides $g_{\mu 
\nu}$): two scalars: the dilaton $\varphi = \Phi / \sqrt 2$ and the 
$R-R$ 0-form $C_0$, two 2-forms $B_2 (NS-NS)$ and $C_2 (R-R)$, and 
one ``self-dual'' $R-R$ 4-form $C_4$. The dilaton coupling strengths 
of the forms are: $\lambda_{C_0}^B = 2 \, \sqrt 2$, $\lambda_{B_2}^B 
= - \sqrt 2$, $\lambda_{C_2}^B = + \sqrt 2$ and $\lambda_{C_4}^B = 
0$. [$\lambda_{C_2}^B$ refers to the more complicated mixed 
coupling $e^{\Phi} \, (d \, C_2 - C_0 \, d \, B_2)^2$].
The Kasner exponents 
$p_{\varphi}^B$, $p_i^B$ ($i=1,\ldots ,9$) run over $S_B^8$ defined 
by Eqs.~(\ref{eq3}).

The $D=10$ type I string theory involves (besides $g_{\mu \nu}$ and 
$\varphi$): an ${\rm SO} (32)$ vector potential,
and a 2-form. The
dilaton couplings are $\lambda_1^I = \sqrt 2 / 2$ 
and $\lambda_2^I = + \sqrt 2$\footnote{Note the misprint in 
Eq.~(12.1.34b) of \cite{polchinski}, corrected on the author's web 
page.}. The Kasner exponents $p_{\varphi}^I$, $p_i^I$ ($i=1,\ldots 
,9$) run, as for IIA and IIB, on the $S_I^8$ defined by 
Eqs.~(\ref{eq3}). 

Finally, the $D=10$ heterotic string theories involve (besides 
$g_{\mu \nu}$ and $\varphi$): an ${\rm SO} (32)$ or $E_8 \times E_8$ 
vector potential, and a 2-form. Their respective (Einstein-frame) 
dilaton couplings are: $\lambda_1^h = - \sqrt 2 / 2$, $\lambda_2^h = 
- \sqrt 2$. The Kasner sphere $S_h^8$ for $p_{\varphi}^h$, $p_i^h$ 
($i=1,\ldots ,9$) is the same as for IIA, IIB or I. 

Let us denote for each given theory ``th'' (where ${\rm th} = 
M,A,B,I,h$ labels the theory) the full (finite) sequence of stability 
exponents as $w_{J}^{\rm th} (p)$, where $J$ labels all the possible 
exponents within each theory. E.g., when ${\rm th} = M$ the label $J$ 
takes 690 values corresponding to 
the set $\{ w_J^M \} = \{ g_{\alpha \beta 
\gamma}^M , e_{\alpha_1 \alpha_2 \alpha_3}^{M(3)} , b_{\beta_1 \ldots 
\beta_6}^{M(3)} \}$.
The condition of ``Kasner 
stability'' of each theory is that there exist an open region of the 
corresponding Kasner sphere $S_{\rm th}^8$ where $w_J^{\rm th} (p) > 
0$ for {\it all} the labels $J$. 
However, we have proven that, for all theories, 
${\rm inf}_J \, w_J^{\rm th} (p)$ is strictly 
negative for all values of $p \in S_{\rm th}^8$, except at a finite 
number of isolated points where it vanishes.

Let us first consider $M$-theory.
We have proven a stronger result, namely, that the electric
stability conditions alone are never fulfilled.
If, at any point on $S_M^8$, we order the 
Kasner exponents as $p_1^M \leq p_2^M \leq \cdots \leq p_{10}^M$, the 
most stringent electric stability criterion involves
$f_0(p) \equiv  p_1^M + p_2^M + p_3^M$. 
To show that this function is non-positive on
the cell $p_1^M  \leq \cdots \leq p_{10}^M$
of the Kasner sphere, we maximize it
subject to the 
constraints $\sum \, (p_{\alpha}^M)^2 = 1$, $\sum \, p_{\alpha}^M = 
1$.
These constraints can be taken into account by introducing two 
Lagrange multipliers. 
After a straightforward (but rather long) 
exhaustive analysis,
we have found that 
$f_0^{\max} = 0$, this maximum being reached only at $p_1 = \cdots = p_9 = 
0$, $p_{10} = 1$.

To deal with the type IIA theory, we use the fact that IIA is the 
Kaluza-Klein (KK) reduction of $M$ on a circle. This fact dictates 
the link between the field variables of the two models. If we label 
by the letter $y$ the compactified dimension this link 
implies the following relation between the 
(Einstein-frame) Kasner exponents of the two theories ($i=1, \ldots 
,9$)
\begin{equation}
p_{\varphi}^A = \frac{6 \, \sqrt 2 \, p_y^M}{8 + p_y^M} \quad , \quad 
p_i^A = \frac{8 \, p_i^M + p_y^M}{8 + p_y^M} \, . \label{eq15}
\end{equation}
Forgetting about this Kaluza-Klein motivation\footnote{Note that the 
fact that the IIA field variables depend on one less variable than 
the $M$-ones is unimportant. What is important is the map 
(\ref{eq15}) and the fact that we have taken into account in the
stability criteria 
all possible dangerous terms in a generic solution.}, 
we can consider that Eqs.~(\ref{eq15}) define a one-to-one map 
$\pi_{AM}$ from $S_M^8$ to $S_A^8$: $p_{\alpha}^A = \pi_{AM} \, 
(p_{\beta}^M)$. Using this map, we have then shown that the complete 
set of IIA stability conditions is 
logically equivalent to the complete set of 
$M$ stability conditions.
The instability of the 
Kasner behaviour of $M$-theory proven above then implies 
that the Kasner behaviour of IIA is also unstable.

To deal with the type IIB theory, we use 
the fact that IIA and IIB 
are related by $T$-duality. The link between the field 
variables of the two models dictated by $T$-duality \cite{BHO} 
enables one to 
derive a certain fractionally linear map $\pi_{BA}$
between their (Einstein-frame) Kasner 
exponents, which can be used, as above, to prove the
Kasner-stability equivalence of the types IIA and IIB theories.
Since type IIA is unstable, type IIB is also unstable.

At this stage, we know that $M$, IIA and IIB are {\it equivalent} 
with respect to Kasner stability, and are all unstable. It remains to 
tackle the type I and heterotic theories, which 
are equivalent because their stability conditions 
are algebraically mapped onto each other by the $S$-duality
transformation
$p_{\varphi}^I = - p_{\varphi}^h \quad , \quad p_i^I = p_i^h $. 
To study the Kasner-stability of the heterotic theory, we found 
very convenient to replace the Einstein-frame Kasner 
exponents $(p_{\varphi}^h , p_i^h)$ by their {\it string-frame} 
counterparts $(\alpha_i^h)$. The link between the two is (in $d+1$ 
spacetime dimensions, see, e.g. \cite{BDV})
\begin{equation}
p_{\varphi} = \frac{\sqrt{d-1} \, \sigma}{d-1-\sigma} \quad , \quad 
p_i = \frac{(d-1) \, \alpha_i - \sigma}{d-1-\sigma} \, , \label{eq18}
\end{equation}
with $\sigma \equiv \left( \sum_i \, \alpha_i \right) - 1$ and $i = 
1,\ldots ,d$. In terms of the $\alpha$'s the Kasner sphere $S^{d-1}$ 
is simply the usual unit sphere, $\sum_i \, (\alpha_i)^2 = 1$. In our 
case, $d=9$ and one should add a label ``$h$'' to both the $p$'s and 
the $\alpha$'s. In terms of the string-frame exponents it is found 
that the $h$-stability conditions
are equivalent to the simpler 
inequalities $\alpha_i^h > 0$ and $\alpha_i^h + \alpha_j^h +
\alpha_k^h < 1$ 
(where $i,j,k$ are all 
different) subjected to the constraints 
$\sum_i \, (\alpha_i^h)^2 = 1$.
It is easy to verify that these inequalities can never hold when the space 
dimension is $d=9$. In that case, the closest one comes to satisfying 
the inequalities is the isotropic point
$\alpha_i = 1/3$ for which the second 
inequality is saturated. 
This concludes our 
proof that the heterotic model (and therefore also the type I one) is 
Kasner unstable. Finally the two blocks of theories $(M,A,B)$ and 
$(I,h)$ are both Kasner unstable, though for different algebraic 
reasons.

Our results so far show that the generic solution of the low-energy 
string models can never reach a monotonic Kasner-like behaviour. 
Following the BKL approach \cite{BKL} one can go further and study 
the evolution near a cosmological singularity as a sequence of 
Kasner-like ``free flights'' interrupted by ``collisions'' against 
the ``potential walls'' corresponding to the various 
stability-violating exponents $g$, $e$ or $b$. We have studied this
problem \cite{DH00} and found the following universal
``collision law'' giving the Kasner exponents
$\bar{p}'^{\mu}$ of the Kasner epoch following a collision
in terms of the old ones:
\begin{equation}
\bar{p}'^{\mu} = \left( 1-2 \ \frac{(w \cdot \bar{p}) \, 
(w \cdot u)}{(w \cdot 
w)} \right)^{-1} \left[ \bar{p}^{\mu} - 2 \ \frac{(w \cdot \bar{p}) \, 
w^{\mu}}{(w \cdot w)} \right] \, . \label{eq27}
\end{equation}
Here, $\bar{p}^{\mu}$ stands for $\bar{p}^0 \equiv p_\varphi$ and
$\bar{p}^i \equiv p_i$.  The scalar products are computed with the
metric $G_{\mu \nu}$ occurring in the quadratic form (\ref{eq3a}),
namely, $G_{00} = 1$, $G_{0i} = 0$, $G_{ij} = \delta_{ij} - 1$, while
the vector $u$ (entering (\ref{eq3b}))
has ``covariant'' components $u_0 = 0$, $u_i = 1$.
Finally, the ``contravariant'' vector $w^\mu$ 
characterizes the ``wall'' responsible for the
collision and is defined in such a way
that the corresponding exponent ($g(p)$, $e(p)$ or $b(p)$) 
reads $w(p) = w_\mu \bar{p}^\mu \equiv G_{\mu \nu} w^\mu \bar{p}^\nu$. 
[E.g., 
for the wall associated with the
electric exponent $e_{123}(p) \equiv p_1 + p_2 + p_3 $,
$w_\mu$
reads $w_0 = 0$, $w_i = 1$ for $i = 1,2,3$ and $w_i = 0$ for
$i>3$.]
The result (\ref{eq27}) (which is a rescaled
geometrical reflection in the hyperplane
$w_\mu \bar{p}^\mu = 0$)
applies uniformly to all possible ``walls'': 
gravitational, electric or magnetic. It generalizes
particular results derived by many authors \cite{collision}.

Summarizing: In all string models, the generic solution near a 
cosmological singularity for the massless bosonic degrees of freedom 
exhibits BKL-type oscillations, i.e. 
a (formally infinite)
alternation of Kasner-epochs. The primary sources of this
BKL behaviour are 
(i) the presence of $p$-forms in the field spectrum of the theories
and, (ii) the strength of their dilaton couplings.  In the
absence of $p$-forms, or if the $\lambda_p$'s were somewhat
smaller, the monotonic Kasner
behaviour would be stable and generic. 
The general rule defining the change 
of Kasner exponents from one epoch 
to the next is given by 
Eq.~(\ref{eq27}), where $w$ is the ``wall'' (among the various 
gravitational, electric or magnetic ones) for which $w(p) = w_\mu
\bar{p}^\mu$ is most 
negative. We anticipate that the discrete dynamics (\ref{eq27}) will 
define (in all string models) a chaotic motion on the Kasner sphere. 
At this stage, the physical consequences of such a chaotic motion are 
unclear. It might constitute a problem for the pre-big-bang scenario 
\cite{PBB} which strongly relies on the existence, near a (future) 
cosmological singularity, of relatively large, quasi-uniform patches 
of space following a monotonic, dilaton-driven Kasner behaviour. By 
contrast our findings suggest that the {\it spatial inhomogeneity} 
continuously {\it increases} toward a singularity, as all 
quasi-uniform patches of space get broken up into smaller and smaller 
ones by the chaotic oscillatory evolution. In other words, the 
spacetime structure tends to develop a kind of ``turbulence''
\cite{turbulence}.

We are aware of the limitations of our result 
(tree-level bosonic massless modes only) but we think that our finding 
suggests that the full quantum, string-theory behaviour might be at 
least as complicated, near a cosmological singularity, as our 
simplified analysis shows.

We thank Volodia Belinskii, Isaak Khalatnikov and Ilan Vardi for
useful exchanges of ideas.
T.D. is grateful to David Gross, Gary Horowitz and Joe Polchinski for 
informative discussions.  M. H. thanks the Institut des Hautes Etudes
Scientifiques for its kind hospitality.

\end{document}